\documentclass[useAMS,usenatbib]{mn2e}

\usepackage{graphicx}
\usepackage{dcolumn}
\usepackage{bm}
\usepackage{color}
\usepackage{natbib}
\usepackage{color}
\usepackage{lscape}

\newcommand{\qm}[1]{``#1''}

\def\sss{\scriptscriptstyle}

\def\K{{\sss \!K}}

\def\P{{\sss \!P}}
\def\S{{\sss \!S}}

\def\nur{\nu_\mathrm{r}}

\def\nuK{\nu_\K}

\def\nuP{\nu_\P}

\def\inc{i}
\def\ex{e}
\def\fn{signal~fraction~}

\definecolor{gray}{rgb}{.6,.6,.6}
\definecolor{green}{rgb}{0,.6,0}
\definecolor{red}{rgb}{.9,0,0}

\title[Orbital motion in strongly curved spacetime: the observable signal]{Power density spectra of modes of orbital motion in strongly curved spacetime: obtaining the observable signal}

\author[P. Bakala et al.]
{P. Bakala$^1$, G. T\"or\"ok$^1$, V. Karas$^2$, M. Dov\v{c}iak$^2$, M. Wildner$^{1,3}$, D. Wzientek$^1$,
\newauthor E. \v{S}r\'{a}mkov\'{a}$^1$, M. Abramowicz$^{1,4,5}$, K. Goluchov\'{a}$^1$, G. P. Mazur$^{5,6}$, F. H. Vincent$^{5,7}$
\\$^1$ Institute of Physics, Faculty of Philosophy and Science, Silesian University in Opava, Bezru\v{c}ovo n\'{a}m. 13, CZ-74601 Opava,\\ Czech Republic
\\$^2$ Astronomical Institute, Bo\v{c}n\'{\i}\ II 1401, CZ-14131 Prague, Czech Republic
\\$^3$ Department of Theoretical Physics and Astrophysics, Faculty of Science, Masaryk University,  CZ-61137 Brno, Czech Republic
\\$^4$ Physics Department, Gothenburg University, SE-412-96 G\"oteborg, Sweden
\\$^5$ Copernicus Astronomical Center, ul. Bartycka 18, PL-00-716 Warszawa, Poland
\\$^6$ Institute for Theoretical Physics, University of Warsaw, Hoza 69, 00-681 Warsaw, Poland
\\$^7$ Laboratoire AstroParticule et Cosmologie, CNRS, Universite Paris Diderot, 10 rue Alice Domon et Leonie Duquet, 75205,\\ Paris Cedex 13, France}

\begin{document}

\date{Accepted 2014 January 10.  Received 2013 November 21; in original form 2013 April 26}

\maketitle

\label{firstpage}

\begin{abstract}
High frequency quasi-periodic oscillations (HF QPOs) appear in the X-ray variability of several accreting low-mass binaries. In a series of works it was suggested that these QPOs may have connection to inhomogeneities orbiting close to an inner edge of the accretion disc. In this paper we explore the appearance of an observable signal generated by small radiating circular hot spots moving along quasi-elliptic trajectories close to the innermost stable circular orbit in the Schwarzschild spacetime. {Our consideration takes into account the capabilities of observatories that have been operating in the past two decades represented by the Rossi X-ray Timing Explorer (RXTE) and the proposed future instruments represented by the Large Observatory for X-ray Timing (LOFT).}
For these purposes we choose such model parameters that lead to lightcurves comparable to those observed in Galactic black hole sources, in particular the microquasar GRS~1915+105.
We find that when a weak signal corresponding to the hot-spot Keplerian frequency is around the limits of the RXTE detectability, the LOFT observations can clearly reveal its first and second harmonics. Moreover, in some specific situations the radial epicyclic frequency of the spot can be detected as well. Finally, we also compare the signal produced by the spots to the signal produced by axisymmetric epicyclic disc-oscillation modes and discuss the key differences that could be identified via the proposed future technology. {We conclude that the ability to recognize the harmonic content of the signal can help to distinguish between the different proposed physical models.}
\end{abstract}

\begin{keywords}
X-rays: binaries -- accretion, accretion disks -- black hole physics - gravitation -- methods: numerical
\end{keywords}

\section{Introduction}
\label{section:intro:QPO}
Several low-mass X-ray binaries (LMXBs) exhibit in the high frequency part of their X-ray power density spectra (PDS) distinct peaks, so-called  quasi-periodic oscillations (QPOs). The oscillations with the highest frequencies are sometimes displayed in the range of hundreds Hz (HF QPOs). In neutron star (NS) sources two simultaneous strong HF QPO peaks are often found (referred to as the upper and the lower QPO). The frequencies of these twin peak QPOs are varying over time, sometimes by several hundreds Hz on timescales of days, and the fastest documented changes of twin QPO frequency are of the order of Hz/sec \citep[][]{pal-etal:2004,bar-etal:2005,bar-vau:2012}. The frequency ratio sampled from the available observations of individual NS QPO sources often clusters (typically around the 3:2 value). The clustering can account either to incomplete data sampling, weakness of the two QPOs outside the limited frequency range, or to the intrinsic source clustering \citep[see][for details]{abr-etal:2003a, bel-etal:2005,bel-etal:2007, tor-etal:2008a, tor-etal:2008b, tor-etal:2008c, tor:2009, bar-bou:2008, bou-etal:2010,wan-etal:2013}. The HF QPOs observed in microquasars exhibit the 3:2 frequency ratio \citep[e.g.,][]{abr-klu:2001,mcc-rem:2006} as well. These QPOs are, however, somewhat different from the case of NS observations. Their frequencies are much more stable while their amplitudes are lower. One should also note that they are not displayed frequently and exhibit spectral differences from the NS QPOs \citep[e.g.][]{Kli:2006:CompStelX-Ray:}.

So far, there is no consensus on the QPO origin, and numerous models have been proposed \citep[e.g.,][and others]{alp-sha:1985,lam-etal:1985,mil-etal:1998,psa-etal:1999,wag:1999,wag-etal:2001,abr-klu:2001,tit-ken:2002,rez-etal:2003,pet:2005,zha:2005,sra-etal:2007,kat:2007,stu-etal:2008,hor-etal:2009,muk:2009,lai-etal:2012}. Several but not all of the models expect that the generic QPO mechanism is the same for both NS an BH sources. Commonly, there is a belief that the HF QPOs carry important information about the inner accreting region dominated by the effects of strong Einstein's gravity. In several works it was suggested that the observed modulation of X-ray flux may have some connection to the inhomogeneities propagating close to the inner edge of the accretion discs. This idea was first discussed within the hypothesis of PDS continuum origin due to the presence of spots on the accretion disk surface \citep[e.g,][]{abr-etal:1991,min-etal:1994}.\footnote{In the same context, \citet{abr-etal:1992} argue that the surfaces of accretion disks must  be populated by vortices. They point out that  in differentially rotating fluids, vortices are not to be quickly destroyed by shear. Instead, they may be a very long lived objects, that survive in the disc like the Red Spot in Jupiter's atmosphere.}

Spots as the origin of QPOs have been discussed, in particular, in a series of papers \citep{ste-vie:1998a,ste-vie:1998b,ste-vie:1999,ste-vie:2002,mor-ste:1999} that explain QPOs as a direct manifestation of modes of relativistic epicyclic motion of blobs at various radii $r$ in the inner parts of the accretion disc. Within the model, the two observed HF QPO peaks arise due to the Keplerian and periastron precession of the relativistic orbits.
In a similar manner, \cite{cad-etal:2008}, \cite{kos-etal:2009}, and \cite{ger-etal:2009} introduced a concept in which the QPOs were generated by a {tidal disruption} (TD) of large accreting inhomogeneities. Recently, \cite{ger:2012} argued that the behaviour of  the azimuth phase $\phi(t)$ for non-closed quasielliptic orbits in the curved spacetime can be responsible for the observed pairs of HF QPOs.

In this paper we investigate the behaviour of the observable signal produced by radiating small circular hotspots. We discuss the detectability of the produced signal propagated from the strongly curved spacetime region. In our discussion we consider the capabilities of the present X-ray observatories represented by Rossi X-ray Timing Explorer (RXTE), as well as the proposed future instruments represented by the Large Observatory for X-ray Timing (LOFT). We also compare the signal produced by spots to the signal obtained from another specific kind of simulations assuming axisymmetric epicyclic disc-oscillation modes.

\section{Tracing the signal from spot}
\label{Section:Spots}

Effects of special and general relativity act on photons emitted by the hot matter that orbits in the inner part of the accretion disc with velocity reaching values of percents of speed of light. Complex fast calculations of the flux propagation through curved spacetime towards a distant observer represent a demanding, although, in principle, well understood problem with practical astrophysical applications. In past decades, serious effort has been invested to solving it using mathematically elegant approaches, as well as straightforward ray-tracing methods profiting from currently available high computational power. Overall, various alternative strategies have been developed by different authors \citep[e.g.][and others]{fel-etal:1974,cun:1975,lao:1991,kar-bao:1992,kar-etal:1992,stu-bao:1992,vie:1993,mat-etal:1993,zak:1994,bro-etal:1997,dab-etal:1997,bur-etal:2004,bak-etal:2007,vin-etal:2012}. Particular application to the timing of radiating spots orbiting around black holes was discussed in greatest detail by \cite{sch-ber:2004a,sch-ber:2004b} who elaborated the model of small radiating spots described as isotropic, monochromatic emitters following non-circular geodesic trajectories and also investigated effects given by large azimuthal shearing of such spots.

The expected spot-signal can be in principle used to model the response of X-ray detectors such as the Proportional Counter Arrray (PCA) on the board of the RXTE \citep[][]{jah-etal:1996,jah-etal:2006}. The actual response of the instrument, however, depends on the complex time-dependent spectral behaviour of the whole source flux. In order to obtain relevant results one must then use a complete physical model of the accreting source radiation. First steps in this direction were taken in the above mentioned work of \citeauthor{sch-ber:2004a} who investigated their hot-spot lightcurves assuming steady state disk models. This option for modelling the observable signal might be, however, difficult. The current observed radiation of Galactic microquasars exhibits a complex behaviour of transitions between particular X-ray states which are yet poorly understood, and the detailed selfconsistent picture of full accreting system physics is rather missing so far. Another option which we follow here is to use an empirical model of the global source flux and add just the particular investigated model for the HF QPO variability including the predicted time dependent energy spectra. This allows us to reproduce well the detectable signal using the response matrices of the RXTE and compare them to the results that can be achieved through the Large Area Detector (LAD) currently proposed within the future LOFT mission \citep[][]{fer-etal:2012}.  

\begin{figure*}
\begin{minipage}{1\hsize}
\begin{center}
\begin{minipage}{1\hsize}
\includegraphics[width=\hsize]{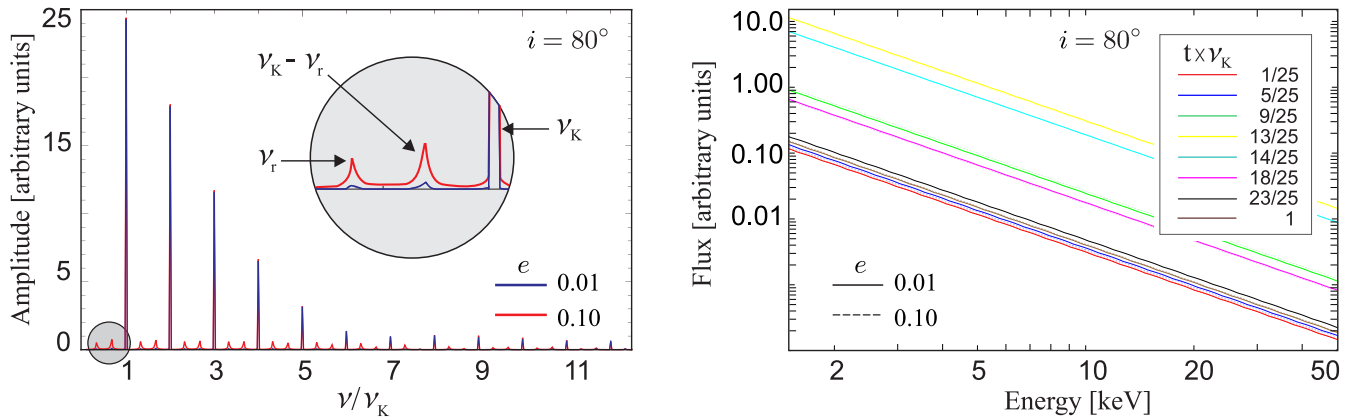}
\end{minipage}
\end{center}
\end{minipage}
\caption{Expected net spot flux measured by a distant observer close to the equatorial plane. {Left:} Amplitude spectrum. {{Right:}} Time dependent energy spectra drawn for a distant observer. }
\label{figure:D80}
\end{figure*}

\subsection{Simulation setup}
\label{section:setup}

Next we investigate the signal from the motion of a small radiating spot obtained using the extended KYspot code \citep{dov-etal:2004}. For simplicity, we choose the zero  black hole spin.  {We assume radial perturbations of the circular orbit of the radiating spot at the specific orbital radius {$r_{3:1}=6.75M$}. At this radius, the radial epicyclic frequency $\nur$ is exactly three times smaller than the Keplerian frequency $\nuK$. The periastron precession frequency $\nuP$ then reaches the value $\nuP=\nuK-\nur=2/3\nuK$.} The intrinsic local spot spectrum is chosen in a power law form,
\begin{eqnarray}
\label{equation:local:spectrum}
&&N(E)= kE^{-1.9}.
\end{eqnarray}
For the further elaboration of observable effects we set the black hole mass $M\doteq11M_\odot$. The considered mass scales the Keplerian frequency of the spot to 160Hz, close to one of HF QPO frequencies observed in the microquasar GRS-1915+105 \citep[e.g.][]{mcc-rem:2006}. {For the spot trajectory we assume {constant angular velocity equal to the Keplerian value $\Omega=\Omega_\mathrm{K}^{3:1}$ and} two different amplitudes $\mathcal{E}$ of the radial epicyclic oscillation - $\mathcal{E}=0.1M$ and $\mathcal{E}=1M$, respectively. These amplitudes correspond to a dimensionless pseudoeccentricity 
\begin{equation}
\label{equation:pseudoeccentricity}
\ex\equiv\frac{(r_\mathrm{max}-r_\mathrm{min})}{(r_\mathrm{max}+r_\mathrm{min})}=\mathcal{E}/r
\end{equation}
reaching the value of $\ex\doteq0.01$ and $\ex\doteq0.1$, respectively.} The integration interval of the spot observation measured by an observer at the infinity is chosen as 10sec. Due to the technical limitations of the applied code we restrict ourselves to this integration time in all further simulations presented in this paper.

We assume the global source flux described by the spectral distribution $N(E)$ and power density spectra $P(\nu)$,
\begin{eqnarray}
\label{equation:global:spectrum}
&&N(E)= kE^{-2.5},\\
\label{equation:global:variability}
&&P(\nu)= p_0\nu^{p_1}+ \frac{1}{\pi}\frac{p_3p_4}{(\nu-p_2)^2+p_3^2},
\end{eqnarray}
where $k$ is chosen to normalize the assumed countrate roughly to $1$Crab and $p_{\mathrm{i}}=[0.001,-1.3,2.5,0.8,0.002]$. This setup mimics the so-called high steep power law (HSPL) state in GRS~1915+105, including steep spectrum and power-law dominated variability with an additional broad Lorentzian component at low frequencies \citep[][]{mcc-rem:2006}. 

\subsection{Signal behaviour}

\begin{figure*}  
\begin{center}
\begin{minipage}{1\hsize}
\includegraphics[width=\hsize]{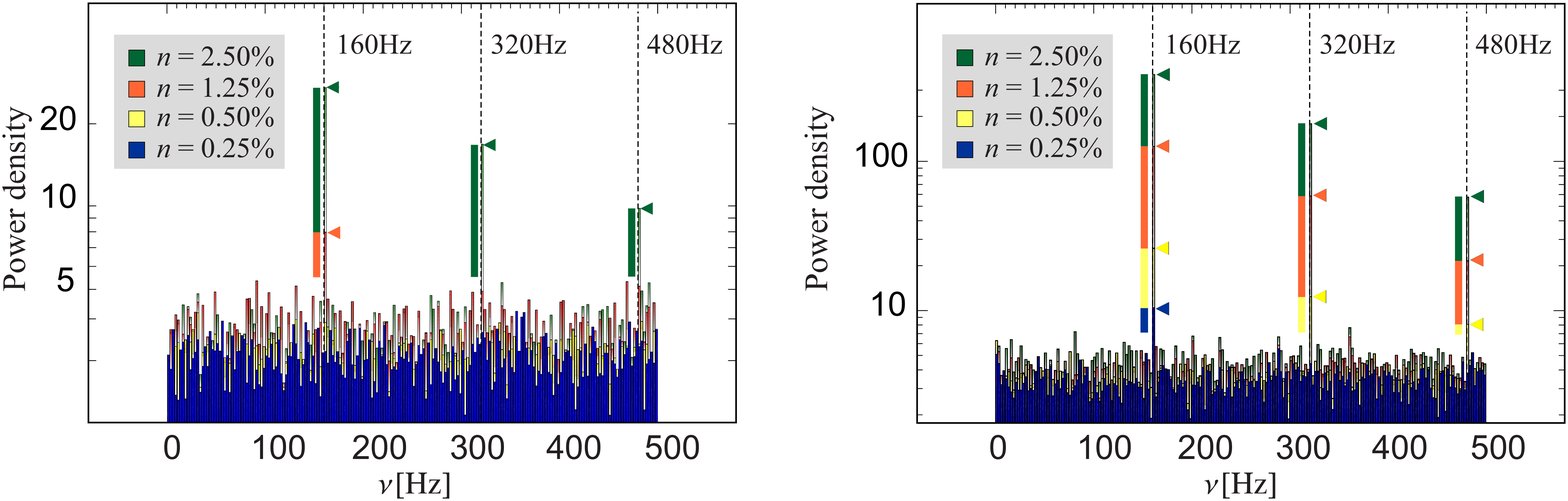}
\end{minipage}
\end{center}
\caption{{Leahy-normalized PDS obtained for various levels of \fn $n$ assuming nearly equatorial view ($\inc=80^{\circ}$) and pseudoeccentricity $\ex=0.1$. The colour-coded flags emphasize the signal excess above the noise level for given $n$.} {Left: Outputs for an exemplary source regardinmg the RXTE capabilities. Right: Outputs for the same source luminosity considering the LOFT capabilities. Note that the lowest displayed values of $n$ illustrated by the blue and yellow colour do not indicate any significant features within the RXTE PDS. On the other hand, the LOFT PDS reveals the Keplerian frequency (blue and yellow PDS), resp. its first two harmonics (yellow PDS).}}
\label{figure:PDSD80}
\end{figure*}

\begin{figure*}
\begin{minipage}{1\hsize}
\begin{center}
\begin{minipage}{1\hsize}
\includegraphics[width=\hsize]{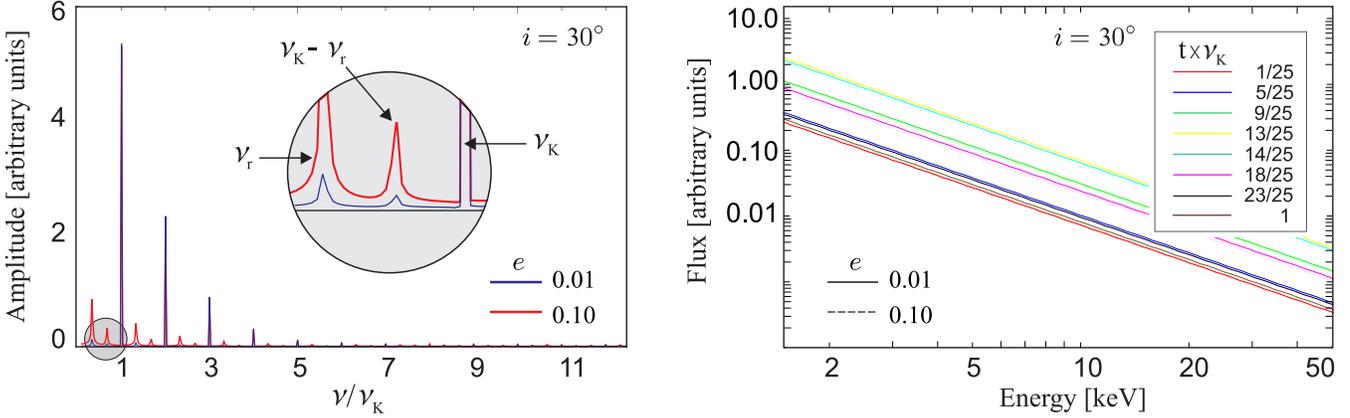}
\end{minipage}
\end{center}
\end{minipage}
\caption{Expected net spot flux measured by a distant observer close to vertical axes. {Left:} Amplitude spectrum. {{Right:}} Time dependent energy spectra drawn for the distant observer.}
\label{figure:D30}
\end{figure*}


Since the signal from the spot strongly depends on the source inclination $\inc$, we compare the results for two representative values of $\inc$ corresponding to the nearly equatorial view and the view close to vertical axes.

\subsubsection{Nearly equatorial view}

In Figure~\ref{figure:D80} we include amplitude spectra and time dependent energy spectra of the net spot signal calculated for the distant observer assuming $\inc=80^{\circ}$. The spot signal is dominated by the Keplerian frequency and its harmonics amplified by relativistic effects which is well illustrated by the amplitude spectrum on the left panel of Figure \ref{figure:D80}. {The eccentricity given by $\ex=0.01$ causes only a negligible modulation at the radial and precession frequency. The increased eccentricity corresponding to $\ex=0.1$  can be well recognized in the amplitude spectra, but the signal is still dominated by the Keplerian frequency and its harmonics.} The time dependent energy spectra of the spot are depicted in the right panel of Figure~\ref{figure:D80}. We can see that they clearly reveal the signatures of relativistic redshift effects.

So far we have reproduced just the variability and spectra of the net spot flux. In order to assess the observable effects we have to study the total composition of the net spot flux together with the global source flux given by Equations (\ref{equation:global:spectrum}) and (\ref{equation:global:variability}). Assuming this composed radiation, we can consider the capabilities of the RXTE and LOFT instruments using their response matrices and provided software tools. The time-dependent spectra describing the composed radiation are then convolved with the appropriate response matrix giving an estimate of the observed data. These are Fourier transformed to the resulting power spectra. The detectability of the spot signatures then depends obviously on the fraction of photons from the spot in the total flux. Hereinafter we refer to this quantity which determines the signal to noise ratio shortly as \fn $n$.

Figure~\ref{figure:PDSD80} shows PDS resulting from the RXTE and LOFT simulations assuming various levels of $n$ and the inclination $\inc=80^{\circ}$ (i.e., nearly equatorial view). It includes the cases when the signal is weak for the RXTE and there are no significant features within its PDS as well as the high \fn when first two harmonics of the Keplerian frequency can be seen. Comparing both panels of this Figure we can deduce that when the weak QPO signal corresponding to the hot-spot Keplerian frequency is around the limits of the RXTE detectability, the LOFT observations will clearly reveal its first and second harmonics. We checked that there is, in practice, no qualitative difference between the cases of {$\ex=0.01$ and $\ex=0.1$}. It is therefore unlikely that the periastron precession or radial epicyclic frequency can be detected in addition to the harmonics when the inclination angle is close to the equatorial plane.

\begin{figure*}  
\begin{minipage}{1\hsize}
\includegraphics[width=\hsize]{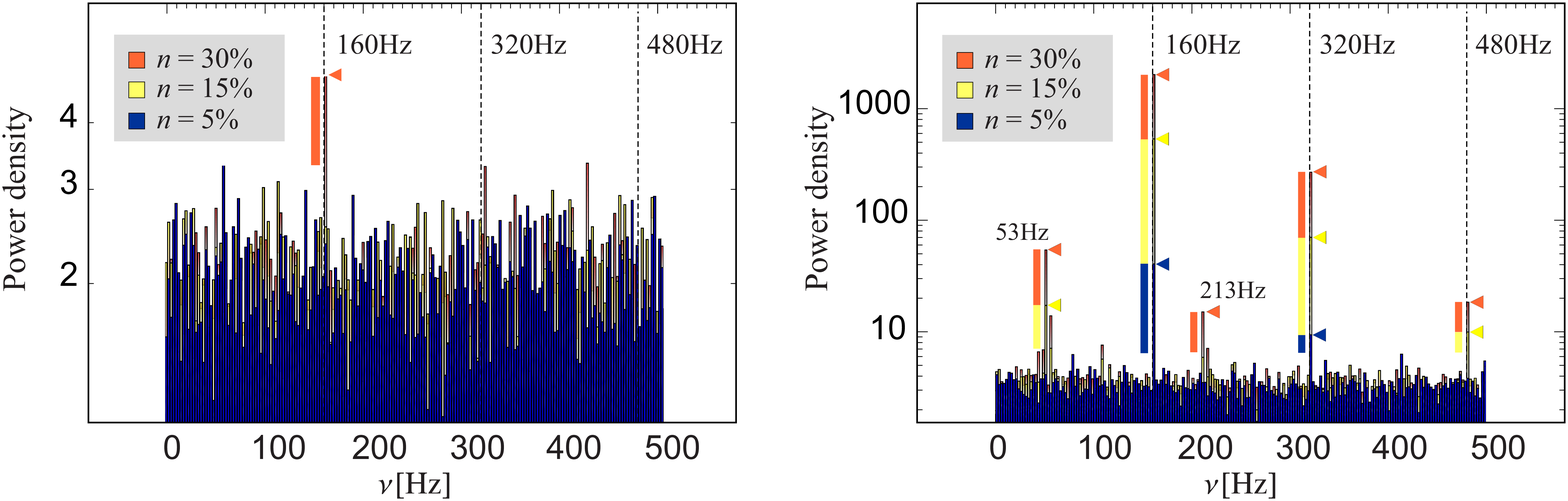}
\end{minipage}
\caption{{Leahy-normalized PDS obtained for various levels of \fn $n$ assuming a view close to the vertical axes ($\inc=30^{\circ}$) and the pseudoeccentricity $\ex=0.1$. The colour-coded flags emphasize the signal excess above the noise level for given $n$. Left: Outputs for an exemplary source, considering the RXTE capabilities. Right: Outputs for the same source luminosity, considering the LOFT capabilities. Note that RXTE PDS includes a barely significant excess of power at 160Hz only for the highest displayed value of \fn $n$. For the same value, the LOFT PDS reveals the first two harmonics and also the radial epicyclic frequency, $\nur\doteq 53$Hz. Moreover, combinational harmonic frequency, $4\nur=2\nuP\doteq213$Hz, can be seen as well. }}
\label{figure:PDSD30}
\end{figure*}

\subsubsection{View close to vertical axes}

For $\inc=30^{\circ}$, the signal is dominated by the Keplerian frequency but the harmonics are much less amplified in comparison to the nearly equatorial view (see the left panel of Figure~\ref{figure:D30}). {The $\ex=0.01$ pseudoeccentricity again causes rather negligible modulation at the radial and the precession frequency. Nevertheless, we can see  that the increased eccentricity of $\ex=0.1$ affects the variability more than for the large inclination angle. Furthermore, the modulation at the radial frequency is comparable to those at the second harmonics of the Keplerian frequency.} 
The time dependent energy spectra are depicted in the right panel of Figure~\ref{figure:D30}.

Figure~\ref{figure:PDSD30} includes PDS resulting from RXTE and LOFT simulations assuming various levels of $n$. It is drawn for {$\ex=0.1$} and includes the few cases when the signal is weak for the RXTE and there are no significant features within its PDS, plus one case when some feature at the Keplerian frequency can be seen. {Comparing both panels of this Figure, one can deduce that when the weak QPO signal corresponding to the hot-spot Keplerian frequency is around the limits of the RXTE detectability, the LOFT observations can clearly reveal its first and second harmonics, and in addition the radial epicyclic frequency $\nur$ and its combinational harmonics $4\nur=2\nuP$. Although it is not directly illustrated in the Figure, it can be shown that decreasing eccentricity to $\ex=0.01$ leads to a similar PDS, only without the features at $\nur$ and $4\nur$.}

{\section{Broad peaks and their harmonics}\label{Section:broad}}

\begin{figure*}  
\begin{minipage}{1\hsize}
\includegraphics[width=\hsize]{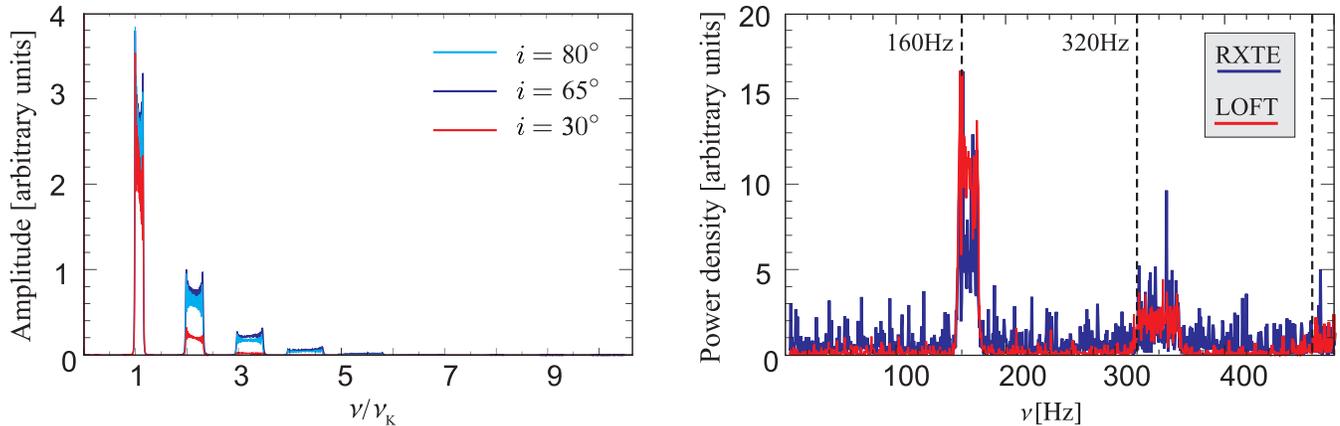}
\end{minipage}
\caption{{Left: Amplitude distribution and its inclination dependence expected for the drifting spot observed from infinity. Right: Illustration of the corresponding PDS drawn for the RXTE and LOFT observations and inclination $\inc=65^{\circ}$.}}
\label{figure:PDSmultiple}
\end{figure*}

In the previous simplified analysis we assumed the single spot orbiting with the fixed Keplerian period. So far, there is no agreement on the physical origin and properties of the hypothetic clumps orbiting in the disc. In various proposed scenarios, the spot signal (de)coherence is commonly related either to the radial distribution of several spots or to the continuous change of the spot orbital radius \cite[see, e.g.,][for details and discussion]{kar:1999,pec-kar:2008,bin-etal:2012}. In the left panel of Figure~\ref{figure:PDSmultiple} we show PDS resulting from consideration of spots created at $r=6.75M$ and drifted to $r=6M$ (ISCO).  The Figure illustrates well the strong inclination dependence of the ratio between the amplitudes observed at Keplerian and harmonic frequencies.  The example of instrument response to the total model flux is illustrated on the right panel of Figure~\ref{figure:PDSmultiple}  (the mass $M$ is set to the same value as in the previous sections).

\begin{figure*}  
\begin{minipage}{1\hsize}
\includegraphics[width=\hsize]{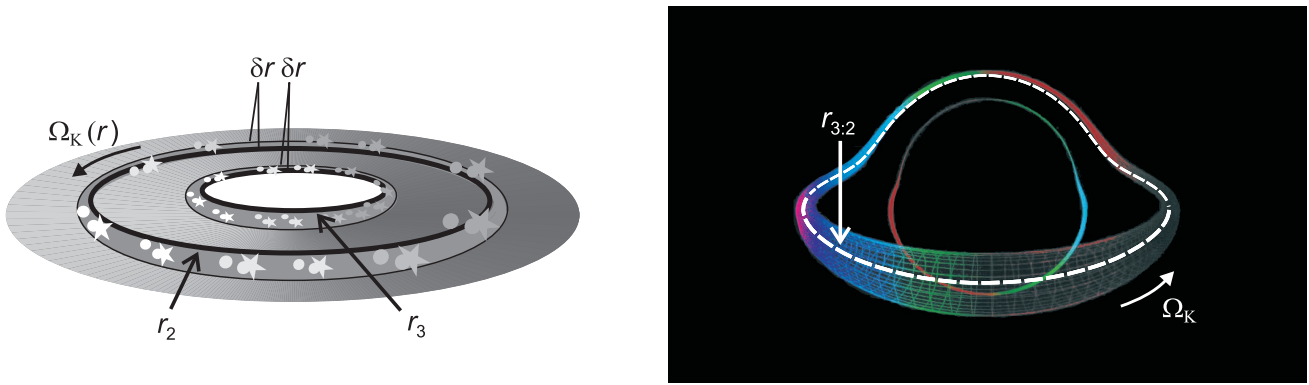}
\end{minipage}
\caption{Left: Sketch of the 3:2 QPO model scheme based on spots orbiting close to two preferred radii. Right: Illustration of the role of lensing and Doppler effects in the visual appearance of torus located at the resonant orbit $r_{3:2}$. }
\label{figure:3to2models}
\end{figure*}

{\subsection{The 3:2 peaks produced by spots and tori\label{Section:comparison}}}

\begin{figure*}  
\begin{minipage}{1\hsize}
\includegraphics[width=\hsize]{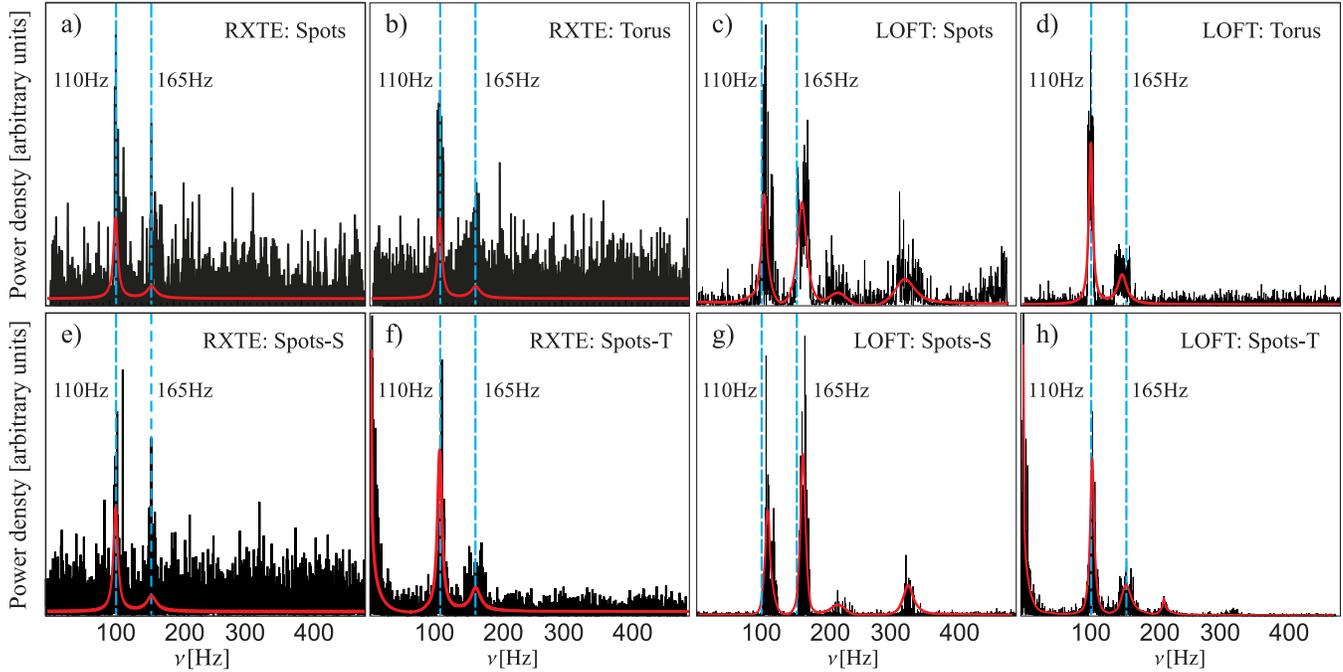}
\end{minipage}
\caption{{a--d) Comparison between the multiple spot and oscillating torus PDS obtained for the two instruments. Superimposed red curves indicate various multi-Lorentzian models. e--h) Results of simulations assuming stationary spots distributed in space (Spots-S) and spots with short lifetimes located just at the two orbits (Spots-T).}}
\label{figure:PDSfit}
\end{figure*}

The radial distribution or drifting of the spots can clearly result in various levels of signal coherence. Nevertheless, small circular spots related to a single preferred radius do not reproduce the often observed 3:2 frequency ratio. Motivated by \cite{kar:1999}, we also consider a somewhat more complicated arbitrary scheme where the multiple spots are created and drifted around radii close to two preferred orbits with Keplerian frequencies roughly in the 3:2 ratio. These orbits are set as $r_{2}=8M$ (radius where the radial epicyclic frequency reaches the maximum value) and $r_{3}=6M$ (ISCO). Spots are then created within the regions $[r_{i}+\delta{r},\,r_{i}]$ with the size given by $\delta{r}=0.75M$. The illustration of this scenario is shown in the left panel of Figure~\ref{figure:3to2models}. Since we keep the same black hole mass as above, $M\doteq11M_\odot$,  our setup leads to the main observable frequencies around 110Hz and 160Hz {(see panels a and c of Figure~\ref{figure:PDSfit} drawn for \fn $n\doteq10\%$ and $\inc=65^{\circ}$)}. In the following consideration we compare the PDS obtained for this setup to the PDS resulting from the model of the oscillating optically thin torus slowly drifting through the resonant radius $r_{3:2}$. 

For the torus kinematics we assume the $m=0$ radial and vertical oscillations with equal intrinsic amplitudes. The possible QPO origin in the resonances between this or similar disc oscillation modes has been extensively discussed in works of \cite{abr-klu:2001}, \cite{abr-etal:2003a,abr-etal:2003b}, \cite{klu-etal:2004}, \cite{hor:2005}, \cite{tor-etal:2005}, and several other authors. Here we adopt the concept previously investigated by \cite{bur-etal:2004} who focused on optically thin torus with slender geometry.

The visual appearance of torus influenced by lensing and Doppler effects is illustrated in the right panel of Figure~\ref{figure:3to2models}.  The periodic changes of the observed luminosity are partially governed by the radial oscillations due to changes of the torus volume while the vertical oscillations modulate the flux just due to lensing effects in the strong gravitational field. The contribution of the two individual oscillations to the variations of the observed flux thus strongly depends on the inclination angle \citep[see also][]{maz-etal:2013}. Here we set $\inc=65^{\circ}$ where the fractions of the power in the two observed peaks are comparable. We set the black hole mass $M=5.65M_{\odot}$ and $a=0$ $(r_{3:2}=10.8M)$, implying that the two oscillatory frequencies are $\nu_\theta(r_{3:2})=160$Hz and $\nu_\mathrm{r}(r_{3:2})\doteq110$Hz. Assuming this setup we produce torus drift lightcurves for the interval $r/r_{3:2}\in[0.97,1.03]$. The resulting PDS drawn for \fn $n\doteq10\%$ is included  in Figure~\ref{figure:PDSfit} {(panels b and d)}. We note that similar PDS can be reached assuming e.g. a near extreme rotating black hole with $a=0.98$ and $M\doteq18M_{\odot}$.  

Using Figure~\ref{figure:PDSfit} we can finally confront the predictions for spots drifted around preferred radii to those expected for the oscillating torus slowly passing the resonant orbit $r_{3:2}$. Inspecting this Figure, we can find that the RXTE PDS obtained for the given setup of the two models are rather similar. {On the other hand, the LOFT PDS reveal the presence/absence of the harmonics in addition to the 3:2 peaks.}

{\subsection{Other examples of the harmonic signature of spot motion}}

{Higher harmonics should appear in the signal generated by small spots in various situations. In next we justify their possible detectability assuming two more examples of the spot motion.}

{Panels e and g of Figure~\ref{figure:PDSfit} marked as \qm{Spots-S} include PDS resulting for the group of 200 stationary spots ($\inc=65^{\circ}$). These spots were spatially distributed around the two orbits $r_2$ and $r_3$ assuming a superposition of Gaussian radial distributions of deviations $\sigma_{2,3}=(0.3,0.1)$ and random orbital phases. In contrary, panels f and h of the same Figure marked as \qm{Spots-T} assume a non-stationary situation with spots that repetitively appear only at the two preferred radii $r_2$ and $r_3$ ($\inc=65^{\circ}$). The spots were created with long delays with respect to their random lifetimes. These were much shorter than the integration time of the simulation. Apparently, in both chosen examples we can find harmonics representing the signature of the spot motion.} 

{Inspecting the Spot-S and Spot-T situations we have found clear harmonics in the expected amplitude spectra for any choice of the model parameters. Nevertheless, we note that their detectability in the observational PDS was not high in all cases. Moreover, as illustrated in the case of Spots-T, various models can produce strong noise at low frequencies.} 

{Detailed elaboration of the question which setup guarantess the detectability of the harmonics thus requires a full study of the related (rich) parameter space. We do not intend to make such a study in this work, but it should be discussed elsewhere.}

\bigskip

\section{Discussion and Conclusions}
\label{section:conclusions}

{We identified the signatures of the spot motion mostly with the harmonic content of the observable signal.} Remarkably, for large inclination angles, the LOFT observations could easily reveal the Keplerian frequency together with its first and second harmonics when the strongest (but still very weak) single signal is around the limits of the RXTE detectability. Nevertheless, the radial epicyclic frequency could be also found providing that the inclination is small. In our analysis we have paid attention to the timing signatures of the motion of small circular spots radiating isotropically from the slightly eccentric geodesic orbits. The case of highly eccentric orbits and/or spot having large azimuthal shear will be presented elsewhere.

{The comparison between spot and torus models in Section~\ref{Section:broad} has been obtained for specific kinematic models. Its general validity is thus limited. For instance, a consideration of resonance driven effects or torus geometrical thickness could also give rise to some harmonic content in the signal from the oscillating tori. Despite these uncertainties, the elaborated comparison indicates clearly that the increased sensitivity of the proposed LOFT mission can be crucial for resolving the QPO nature.}
\vspace{-2ex}

\section*{Acknowledgments}

{We acknowledge the support of the Czech grant GA\v{C}R~209/12/P740, Polish grant NCN UMO-2011/01/B/ST9/05439  and the Swedish VR grant. We further acknowledge the project CZ.1.07/2.3.00/20.0071 "Synergy" aimed to support international collaboration of the Institute of physics in Opava, and the internal SU student grant SGS/11/2013. We thank to the anonymous referee for his comments and suggestions that helped to improve the paper.}


\begin{thebibliography}{99}
%
%

\bibitem[{Abramowicz et~al.} (1991)]{abr-etal:1991} Abramowicz M. A., Bao G., Lanza A., Zhang X.-H., 1991,	A\&A,  245, 454
\bibitem[{Abramowicz et~al.} (1992)]{abr-etal:1992} Abramowicz M. A.,  Lanza A., Spiegel E. A., Szuszkiewicz E., 1992, Nature, 356, 41
\bibitem[{Abramowicz \& Klu{\'z}niak}(2001)]{abr-klu:2001} Abramowicz M. A., Klu{\'z}niak W., 2001, A\&A, 374, L19
\bibitem [{Abramowicz et~al.}(2003a)]{abr-etal:2003a} Abramowicz M. A., Bulik T., Bursa M., Klu{\'z}niak W., 2003a, A\&A, 404, L21
\bibitem [{Abramowicz et~al.}(2003b)]{abr-etal:2003b} Abramowicz M. A., Karas V., Klu{\'z}niak W., Lee W. H., Rebusco P., 2003b, PASJ, 55, 466
\bibitem[{Alpar \& Shaham}(1985)]{alp-sha:1985} Alpar M. A., Shaham J., 1985, Nature, 316, 239 
\bibitem[{Bakala et~al.}(2007)]{bak-etal:2007} Bakala P., \v{C}erm\'{a}k P., Hled\'{i}k S., Stuchl\'{i}k Z., Truparov\'{a} K.,	2007, Cent. European J. of Phys., 5, 599
\bibitem[{Barret et~al.}(2005)]{bar-etal:2005} Barret D., Klu\'zniak W., Olive J. F., Paltani S., Skinner G. K., 2005a, MNRAS, 357, 1288
\bibitem[{Barret \& Vaughan}(2012)]{bar-vau:2012}	Barret D., Vaughan S., 2012, ApJ, 746, 131
\bibitem[{Barret \& Boutelier}(2008)]{bar-bou:2008} Barret D.,  Boutelier M., 2008, New Astron. Rev., 51, 835
\bibitem[{Bao, Hadrava \& Oestgaard}(1994)]{bao-etal:1994} Bao G., Hadrava P., Oestgaard E., 1994, ApJ, 435, 55
\bibitem[{Belloni, M{\'e}ndez  \& Homan }(2005)]{bel-etal:2005} Belloni T., M{\'e}ndez M., Homan J., 2005, A\&A,  437, 209
\bibitem[{Belloni et~al.}(2007)]{bel-etal:2007} Belloni T., Homan J., Motta S., Ratti E., M\'endez M., 2007, MNRAS, 379, 1, 247, e-print arXiv:0705.0793
\bibitem[{Bini et~al.}(2012)]{bin-etal:2012}	Bini D., Falanga M., Geralico A., Stella L., 2012, Class. Quantum Grav., 29, 065014
\bibitem[{Boutelier et~al.}(2009)]{bou-etal:2010} Boutelier M., Barret D., Lin Y., T\"{o}r\"{o}k G., 2010, MNRAS, 401, 1290, e-print arXiv:0909.2990
\bibitem[{Bursa et~al.}(2004)]{bur-etal:2004} Bursa M., Abramowicz M. A., Karas V., Kluzniak W., 2004,	ApJ, 617,  L45
\bibitem[{Bromley, Chen \& Miller }(1997)]{bro-etal:1997} Bromley B. C., Chen K., Miller W. A., 1997, ApJ, 475, 57
\bibitem[{\v{Ca}de\v{z}, Calvani \& Kosti\'{c} }(2008)]{cad-etal:2008} \v{Ca}de\v{z} A., Calvani M., Kosti\'{c}  U., 2008, A\&A, 487, 527
\bibitem[{Cunningham}(1975)]{cun:1975} Cunningham C. T., 1975, ApJ, 202, 788
\bibitem[{Dabrowski et~al.}(1997)]{dab-etal:1997} Dabrowski Y., Fabian A. C., Iwasawa K., Lasenby A. N., Reynolds C. S., 1997, MNRAS, 288, L11
\bibitem[{de Felice, Nobili \& Calvani}(1974)]{fel-etal:1974} de Felice F., Nobili L., Calvani M., 1974, A\&A, 30, 111
\bibitem[{Dov\v{c}iak,  Karas \& Yaqoob}(2004)]{dov-etal:2004}	Dov\v{c}iak M., Karas V., Yaqoob T., 2004, ApJS, 153, 205
\bibitem[{Feroci et~al.}(2012)]{fer-etal:2012} Feroci M. et~al., 2012, Exp. Astron., 34, 415
\bibitem [{Germana et~al.}(2009)]{ger-etal:2009} Germana C., Kosti\v{c} U., \v{C}ade\v{z} A., Calvani M., 2009, in AIP Conf. Proc. Vol. 1126,  SIMBOL-X: Focusing on the Hard X-ray Universe: Proc. 2nd International Simbol-X Symposium, ed. J. Rodriguez \& P. Ferrando (Melville, NY:AIP), p. 367
\bibitem [{Germana}(2012)]{ger:2012} Germana C., 2013, Monthly Notices of the Royal Astronomical Society: Letters, 430, L1
\bibitem[{Hor\'ak et~al.}(2009)]{hor-etal:2009} Hor\'ak J., Abramowicz M. A., Klu{\'z}niak W., Rebusco P., T\"{o}r\"{o}k G., 2009, A\&A, 499, 535
\bibitem[{Hor\'ak }(2005)]{hor:2005} Hor\'ak J., 2005, Astronomische Nachrichten, 326, 824
\bibitem [{Jahoda et~al.}(1996)]{jah-etal:1996} Jahoda K., Swank J. H., Giles A. B., Stark M. J., Strohmayer T., Zhang W., Morgan E. H., 1996,	Proc. SPIE Vol. 2808, p. 59-70, EUV, X-Ray, and Gamma-Ray Instrumentation for Astronomy VII, Oswald H. Siegmund; Mark A. Gummin; Eds.
\bibitem [{Jahoda et~al.}(2006)]{jah-etal:2006} Jahoda K., Markwardt C. B., Radeva Y., Rots A. H., Stark M. J., Swank J. H., Strohmayer T. E., Zhang W., 2006,	ApJS, 163, 401
\bibitem [{Karas \& Bao}(1992)]{kar-bao:1992} Karas V., Bao G., 1992,	A\&A, 257, 531
\bibitem [{Karas, Vokrouhlicky \&  Polnarev }(1992)]{kar-etal:1992}	Karas V., Vokrouhlicky D., Polnarev A. G., 1992, MNRAS, 259, 569
\bibitem [{Karas}(1996)]{kar:1996} Karas V., 1996, ApJ, 470, 743
\bibitem [{Karas}(1999)]{kar:1999} Karas V., 1999, PASJ, 51, 317
\bibitem[{Kato}(2007)]{kat:2007} Kato S., 2007, PASJ, 59, 451
\bibitem[{Klu\'zniak et~al.}(2004)]{klu-etal:2004} Klu\'zniak W., Abramowicz M. A., Kato S., Lee W. H., Stergioulas N., 2004, ApJ, 603, L89
\bibitem[{Kosti\'{c} et~al.}(2009)]{kos-etal:2009} Kosti\'{c} U., \v{C}ade\v{z} A., Calvani M., Gomboc A., 2009, A\&A, 496, 307
\bibitem[{Lai et~al.}(2012)]{lai-etal:2012}	Lai D., Fu W., Tsang D., Horak J., Yu C., 2013, Proceedings of IAUS 290 "Feeding Compact Objects: Accretion on All Scales", C. M. Zhang, T. Belloni, M. Mendez \& S. N. Zhang (eds.), Vol. 290, p. 57, e-print arXiv:1212.5323
\bibitem[{Lamb et~al.}(1985)]{lam-etal:1985} Lamb F. K., Shibazaki N., Alpar M. A., Shaham J., 1985, Nature, 317, 681
\bibitem[{Laor}(1991)]{lao:1991} Laor A., 1991, ApJ, 376, 90
\bibitem[{Miller, Lamb \& Psaltis }(1998)]{mil-etal:1998} Miller M. C., Lamb F. K., Psaltis D., 1998, ApJ, 508, 791
\bibitem[{Mineshige, Takeuchi \&  Nishimori}(1994)]{min-etal:1994}	Mineshige S., Takeuchi M., Nishimori H.,	1994, ApJ,  435, L125
\bibitem[{Morsink \& Stella}(1999)]{mor-ste:1999} Morsink S. M.,  Stella L., 1999, ApJ, 513, 827
\bibitem[{Matt, Fabian \& Ross}(1993)]{mat-etal:1993} Matt G., Fabian A. C., Ross R. R., 1993, MNRAS, 264, 839
\bibitem[{Mazur et~al.}(2013)]{maz-etal:2013}	Mazur G. P., Vincent F. H., Johansson M., \v{S}r\'{a}mkov\'{a} E., T\"{o}r\"{o}k G., Bakala P., Abramowicz M. A., 2013, A\&A, 554, A57, e-print arXiv:1303.3834
\bibitem[{Mukhopadhyay}(2009)]{muk:2009} Mukhopadhyay B., 2009, ApJ, 694, 387
\bibitem[{McClintock \& Remillard}(2006)] {mcc-rem:2006} McClintock J. E, Remillard R. A., 2006, astro-ph/0306213
\bibitem[{Paltani et~al.}(2004)]{pal-etal:2004} Paltani S., Barret D., Olive J. F.,  Skinner G. K., SF2A-2004: Semaine de l'Astrophysique Francaise, meeting held in Paris, France, June 14-18, 2004. Edited by F. Combes, D. Barret, T. Contini, F. Meynadier and L. Pagani. Published by EdP-Sciences, Conference Series, 2004, p. 381
\bibitem[{Pech\'{a}\v{c}ek, Karas \& Czerny}(2008)]{pec-kar:2008}	Pech\'{a}\v{c}ek T., Karas V., Czerny B., 2008, A\&A, 487, 815 
\bibitem[{P\'etri}(2005)]{pet:2005} P\'etri J., 2005, A\&A, 439, L27
\bibitem[{Psaltis et al.}(1999)]{psa-etal:1999} Psaltis D. et~al., 1999,	ApJ, 520, 763
\bibitem[{Rezzolla, Yoshida \& Zanotti  }(2003)]{rez-etal:2003} Rezzolla L., Yoshida S., Zanotti O., 2003,	MNRAS, 344, 978
\bibitem[{Schnittman \& Bertschinger}(2004a)]{sch-ber:2004a} Schnittman J. D., Bertschinger E.,	2004a, ApJ, 606, 1098
\bibitem[{Schnittman \& Bertschinger}(2004b)]{sch-ber:2004b} Schnittman J. D., Bertschinger E.,	2004b,	X-RAY TIMING 2003: Rossie and Beyond. AIP Conference Proceedings, Vol. 714, p. 40
\bibitem[{\v{S}r\'{a}mkov\'{a}, Torkelsson  \& Abramowicz }(2007)]{sra-etal:2007}	\v{S}r\'{a}mkov\'{a} E., Torkelsson U., Abramowicz M. A., 2007, A\&A, 467, 641
\bibitem[{Stella \& Vietri}(1998a)]{ste-vie:1998a}
Stella L.,   Vietri M., 1998a, in Abstracts of the 19th Texas Symposium on Relativistic Astrophysics and Cosmology, ed. J. Paul, T.
Montmerle, \& E. Aubourg (Saclay, France:CEA) 
\bibitem[{Stella \& Vietri}(1998b)]{ste-vie:1998b} Stella L.,  Vietri M., 1998b, ApJ, 492, L59
\bibitem[{Stella \& Vietri}(1999)]{ste-vie:1999} Stella L., Vietri M., 1999, Phys. Rev. Lett., 82, 17
\bibitem[{Stella \& Vietri}(2002)]{ste-vie:2002} Stella L.,  Vietri M., 2002, in The Ninth Marcel Grossmann Meeting, Proc. MGIXMM Meeting held at The University of Rome "La Sapienza",  2000 July  2-8, ed. V. G. Gurzadyan, R. T. Jantzen, \& R. Ruffini, Part A, p. 426
\bibitem[{Stella, Vietri \& Morsink}(1999)]{ste-etal:1999} Stella L., Vietri M., Morsink S. M., 1999, ApJ,  524, L63
\bibitem[{Stuchl\'{\i}k \& Bao}(1992)]{stu-bao:1992} Stuchl\'{\i}k Z., Bao G., 1992, Gen. Relativ. and Gravitation, 24, 945
\bibitem[{Stuchl\'{\i}k et~al.}(2008)]{stu-etal:2008} Stuchl\'{\i}k Z., Konar S., Miller J. C., Hled\'{\i}k S., 2008, A\&A, 489, 963
\bibitem[{Titarchuk \& Wood}(2002)] {tit-ken:2002}	Titarchuk L.,  Wood K.,	2002, ApJ,  577, L23
\bibitem[{T\"or\"ok et~al.}(2005)]{tor-etal:2005} T\"or\"ok G., Abramowicz M.A., Klu{\'z}niak W., Stuchl{\'{\i}}k Z., 2005, A\&A, 436, 1
\bibitem[{T\"{o}r\"{o}k et~al.}(2008a)]{tor-etal:2008a} T{\"{o}}r{\"{o}}k G., Bakala P., Stuchl{\'{\i}}k Z., \v{C}ech P., 2008a, Acta Astronomica, 58, 1
\bibitem[{T\"{o}r\"{o}k et~al.}(2008b)]{tor-etal:2008b} T{\"{o}}r{\"{o}}k G., Abramowicz M. A., Bakala P., Bursa M., Hor\'{a}k J., Klu{\'z}niak W., Rebusco P., Stuchl\'{\i}k Z., 2008b, Acta Astronomica, 58, 15, e-print arXiv:0802.4070
\bibitem[{T\"{o}r\"{o}k et~al.}(2008c)]{tor-etal:2008c} T{\"{o}}r{\"{o}}k G., Abramowicz M. A., Bakala P., Bursa M., Hor\'{a}k J., Rebusco P., Stuchl\'{\i}k Z., 2008c, Acta Astronomica, 58, 113, e-print arXiv:0802.4026 
\bibitem[{T\"{o}r\"{o}k}(2009)]{tor:2009} T{\"{o}}r{\"{o}}k G., 2009, A\&A, 497, 661
\bibitem[{van~der Klis}(2006)]{Kli:2006:CompStelX-Ray:} van~der Klis M., 2006, in {Compact Stellar X-Ray Sources}, ed. W.~H.~G. Lewin    \& M.~van~der Klis (Cambridge: Cambridge Univ. Press),  39, see also astro-ph/0410551
\bibitem[{Viergutz}(1993)]{vie:1993} Viergutz S. U., 1993, A\&A, 272, 355
\bibitem[{Vincent, Gourgoulhon \& Novak}(2012)]{vin-etal:2012}	Vincent F. H., Gourgoulhon E., Novak J.,	2012, Class. Quantum Grav., 29, 245005
\bibitem[{Wang et~al.}(2013)]{wan-etal:2013}	Wang D. H., Chen L., Zhang C. M., Lei Y. J., Qu J. L., 2013, e-print arXiv:1301.5478
\bibitem[{Wagoner}(1999)]{wag:1999}	Wagoner R. V.,	1999, Phys. Rep., 311,  259 
\bibitem[{Wagoner, Silbergleit \&  Ortega-Rodríguez}(2001)]{wag-etal:2001}	Wagoner R. V., Silbergleit A. S., Ortega-Rodríguez M., 2001,	ApJ,  559, L25 
\bibitem[{Zakharov}(1994)]{zak:1994} Zakharov A. F., 1994, MNRAS, 269, 283
\bibitem[{Zhang}(2005)]{zha:2005} Zhang, C. M., 2005, Chin. J.  Astron.  Astrophys.,  5, 21
\end{thebibliography}
\end{document}